\documentclass[reprint,amsmath,amssymb,superscriptaddress,longbibliography]{revtex4-2}

\usepackage{physics}
\usepackage[T1]{fontenc}
\usepackage[utf8]{inputenc}
\usepackage{multirow}
\usepackage{amsmath}
\usepackage{mathtools}
\usepackage{graphicx}
\usepackage{bm}
\usepackage{braket}
\usepackage{soul} 
\usepackage{tcolorbox}
\usepackage{url}
\usepackage[linktocpage, colorlinks=true, linkcolor=darkblue, citecolor=darkblue, urlcolor=darkblue]{hyperref}
\definecolor{darkblue}{rgb}{0,0,.5}
\usepackage{setspace}
\usepackage{booktabs}
\usepackage{algorithm}
\usepackage{algpseudocode}

\usepackage{tabularx}

\begin{document}
\title[]{Ab Initio Many Body Quantum Embedding and Local Correlation in Crystalline Materials using Interpolative Separable Density Fitting}
\author{Junjie Yang}
\affiliation{Division of Chemistry and Chemical Engineering, California Institute of Technology, Pasadena, California 91125, USA}
\affiliation{Marcus Center for Theoretical Chemistry, California Institute of Technology, Pasadena, California 91125, USA}
\author{Ning Zhang}
\affiliation{Division of Chemistry and Chemical Engineering, California Institute of Technology, Pasadena, California 91125, USA}
\affiliation{Qingdao Institute for Theoretical and Computational Sciences and Center for Optics Research and Engineering, Shandong University, Qingdao, Shandong 266237, China}
\author{Shunyue Yuan}
\affiliation{Marcus Center for Theoretical Chemistry, California Institute of Technology, Pasadena, California 91125, USA}
\affiliation{Division of Engineering and Applied Sciences, California Institute of Technology, Pasadena, California 91125, USA}
\author{Jincheng Yu}
\affiliation{Department of Chemistry and Biochemistry, University of Maryland, College Park, Maryland, 20742, USA}
\author{Hong-Zhou Ye}
\affiliation{Department of Chemistry and Biochemistry, University of Maryland, College Park, Maryland, 20742, USA}
\affiliation{Institute for Physical Science and Technology, University of Maryland, College Park, Maryland, 20742, USA}
\author{Garnet Chan}
\email{gkc1000@gmail.com}
\affiliation{Division of Chemistry and Chemical Engineering, California Institute of Technology, Pasadena, California 91125, USA}
\affiliation{Marcus Center for Theoretical Chemistry, California Institute of Technology, Pasadena, California 91125, USA}

\begin{abstract}
    We present an efficient implementation of ab initio many-body quantum embedding and local correlation methods for infinite periodic systems through 
    {translational symmetry adapted} interpolative separable density fitting, an  
     approach which reduces the scaling of the calculations to only linear with the number of k-points.
    Employing this methodology, we compute correlated ground-state coupled cluster energies within density matrix embedding and local natural orbital correlation frameworks for both weakly and strongly correlated solids, using up to 1000 k-points. By extrapolating the local correlation domains and k-point sampling we further obtain estimates of the full coupled cluster with singles, doubles, and perturbative triples ground-state energies in the thermodynamic limit.
\end{abstract}

\maketitle

\raggedbottom
\section{Introduction}
\label{sec:Introduction}


Correlated wavefunction methods have attracted increasing attention in computational materials science,  because they provide a pathway to controllable accuracy via systematically improvable many-body expansions~\cite{marsman-secondorder-2009,muller-wavefunctionbased-2012,booth-exact-2013,yang-initio-2014,kubas-surface-2016,mcclain-gaussianbased-2017,gruber-applying-2018,zhang-coupled-2019,sauer-initio-2019,dovesi-crystal-2020,shi-manybody-2023}.
However, despite their theoretical advantages, these approaches naively suffer from steep computational scaling with system size, which presents an obstacle to obtaining thermodynamic limit (TDL) results~\cite{raghavachari-fifthorder-1989,crawford-introduction-2000,bartlett-coupledcluster-2007,gruber-applying-2018}. 

Consequently, several strategies are commonly used to reduce the cost of such calculations. 
The first exploits the translational symmetry inherent in periodic systems, by utilizing Bloch-type single-particle orbitals labelled by the crystal momentum k.  
When a uniform k-point mesh of size $N_{\text{k}}$ {(containing the $\Gamma$ point)} is employed to sample the Brillouin zone, the calculation is equivalent to one from a periodic supercell containing $N_{\text{k}}$ unit cell images, but with a $N_\text{k}^{-1}$~\cite{motta2019hamiltonian} to 
$N_\text{k}^{-2}$~\cite{mcclain-gaussianbased-2017} reduction in cost.


The second strategy exploits the locality and decay of correlations~\cite{pulay-localizability-1983,kohn-density-1996,lipkowitz-reviews-2007}. This ``near-sightedness'' is embodied in different quantum embedding and local correlation frameworks.
While there are many variations~\cite{li-linear-2002,li-efficient-2006,rolik-generalorder-2011,rolik-efficient-2013,li-clusterinmolecule-2016,hampel-local-1996,hattig-local-2012,riplinger-efficient-2013,schwilk-scalable-2017,guo-communication-2018,ziolkowski-linear-2010,kristensen-mp2-2012,ayala-linear-1999,scuseria-linear-1999,wesolowski-frozen-1993,goodpaster-exact-2010,huang-quantum-2011,libisch-embedded-2014,ye-bootstrap-2020,nusspickel-systematic-2022,shee-static-2024,rebolini-divide-2018,wang-clusterinmolecule-2019,lee-projectionbased-2019,zhu-efficient-2020,wang-clusterinmolecule-2022,chen-xopbc-2020,meitei-periodic-2023,li-clusterinmolecule-2023,ye-periodic-2024,cui-groundstate-2020,cui-efficient-2020,cui-systematic-2022,cui-initio-2025,zgid-dynamical-2011,sun-quantum-2016}, a common feature  is that the system is divided into multiple fragments (or clusters), with each fragment subsequently treated independently or with a simplified level of coupling, for example, via the self-consistency of a single-particle quantity in quantum embedding methods~\cite{sun-quantum-2016}, or through the simplified treatment of weak pairs in certain local correlation methods.
In this work, for concreteness we will focus on density matrix embedding~\cite{knizia2013density,wouters2016practical} as an example of a quantum embedding method, and the local natural orbital (LNO) method~\cite{rolik-efficient-2013,ye-periodic-2024} as an example of a local correlation framework. Both have been applied with success to crystalline materials in a diverse range of applications~\cite{ye-periodic-2024,haldar-local-2023,lau-optical-2024,pham-periodic-2020,lau-regional-2021,cui-efficient-2020,cui-systematic-2022,cui-initio-2025}.


The third strategy uses low-rank representations. Low-rank approximations of the Coulomb tensor have been extensively used, such as Cholesky decomposition and resolution-of-the-identity techniques~\cite{baerends-selfconsistent-1973,whitten-coulombic-1973,jafri-electron-1974,beebe-simplifications-1977,roeggen-beebelinderberg-1986,ren-resolutionofidentity-2012}, pseudospectral approaches~\cite{friesner-solution-1985,neese-efficient-2009,izsak-overlap-2011}, as well as tensor hypercontraction (THC)~\cite{hohenstein-tensor-2012,parrish-tensor-2012,song-atomic-2016,
matthews-critical-2021}.
Interpolative separable density fitting (ISDF) has attracted attention as a protocol for generating THC-like factorizations in systems under periodic boundary conditions, where it has demonstrated the potential to  accelerate multiple electronic structure methods~\cite{hu-interpolative-2017,dong-interpolative-2018,qin-interpolative-2020,qin-machine-2020,sharma-fast-2022,smyser-use-2024,zhang-machine-2024,lu-cubic-2017,duchemin-separable-2019,zhang-machine-2024,gao-accelerating-2020,ma-realizing-2021,duchemin-cubicscaling-2021,malone-overcoming-2019}.

In this work, we describe an implementation of quantum embedding and local correlation for infinite periodic systems that combines the above reduced-scaling strategies. In particular, our methodology exploits translational symmetry through a tensor hypercontraction representation of the Coulomb interaction, implemented via a symmetry adapted ISDF~\cite{yeh-lowscaling-2023,yeh-lowscaling-2024,wu-lowrank-2022,qin-interpolative-2023,rettig-even-2023} based on crystalline orbitals and utilizing the fast Fourier transform (FFTISDF). The primary algorithmic outcome is the ability to perform many-body quantum embedding and local correlation calculations in materials with only linear scaling with the number of k-points. This enables the practical extraction of thermodynamic limit results.

We organize the rest of the work as follows. 
Section~\ref{sec:theory} {recalls} the FFTISDF algorithm for periodic solids utilizing k-point symmetry, and then describes the efficient implementation of the various components of the density matrix embedding and local natural orbital correlation frameworks in this representation and summarizes the parameters used for the benchmark calculations.
Section~\ref{sec:results} provides a set of benchmark studies using coupled cluster theory solvers on representative crystalline systems, including diamond, carbon dioxide, nickel monoxide, and the infinite layer cuprate CaCuO$_2$. These calculations employ Gaussian polarized double-zeta basis sets with up to 1000 k-points to demonstrate the method's scalability, and we extract the thermodynamic limit of the coupled cluster energies for several of the systems.
Finally, Section~\ref{sec:conclusion} summarizes our main findings and discusses potential directions for future work.

\section{Theory}
\label{sec:theory}
Our periodic calculations are performed using a crystalline Gaussian-type orbital(GTO) basis,
\begin{equation}
    \phi_{\mu}^{\boldsymbol{k}}(\boldsymbol{r}) = \sum_{\boldsymbol{T}} 
    \text{e}^{\text{i} \boldsymbol{k} \cdot \boldsymbol{T}} 
    \; \tilde{\phi}_{\mu}(\boldsymbol{r} - \boldsymbol{T})
    \label{eq:periodic-gto}
\end{equation}
where $\tilde{\phi}_{\mu}(\boldsymbol{r})$ is a normal GTO centered on an atom in the reference unit cell; $\boldsymbol{T}$ is a lattice translation vector; and $\boldsymbol{k}$ is a momentum vector sampled in the first Brillouin zone. Our implementation builds upon the Gaussian-Plane-Wave infrastructure provided by the \textsc{PySCF} package~\cite{sun-gaussian-2017,sun-pyscf-2018,sun-recent-2020}, where the basis functions are evaluated on a uniform grid in the unit cell at a set of points $\{ \boldsymbol{r} \}$. (This infrastructure is termed FFTDF in \textsc{PySCF}).

\subsection{Periodic k-point FFTISDF representation of integrals} 
The core idea of the ISDF approach is to approximate products of basis functions on the grid points $\boldsymbol{r}$ by using values from a select set of interpolating points $\boldsymbol{r}_I$ (IPs). This reduces the computational complexity of evaluating two-electron integrals while maintaining systematic improvability of the approximation.
In this work, the interpolation points are selected using a pivoted Cholesky decomposition of {$\phi_{\mu}^{\boldsymbol{k}_\mu}(\boldsymbol{r}) \phi_{\nu}^{\boldsymbol{k}_\nu}(\boldsymbol{r})^*$ (where $\boldsymbol{r}$ is the column index)} closely following the procedure outlined in Ref.~\cite{matthews-improved-2020}. Only the largest $N_{\text{IP}} = c_{\text{IP}} \times N_{\text{AO}}$ columns are used, where the pre-defined constant $c_{\text{IP}}$ controls the accuracy~\cite{lee-systematically-2020}.
This method is suitable for systems with moderate unit cell sizes. 
Although alternative approaches, such as the centroidal Voronoi tessellation k-means algorithm~\cite{dong-interpolative-2018}, may be more efficient for large unit cells, they are not the focus of this study. The interpolation points are used to approximate the product of basis functions on the uniform grid $\{ \boldsymbol{r} \}$:
\begin{equation}
    \begin{aligned}
    \rho_{\mu\nu}^{\boldsymbol{k}_\mu \boldsymbol{k}_\nu}(\boldsymbol{r}) & =
    \phi_{\mu}^{\boldsymbol{k}_\mu}(\boldsymbol{r}) \; \phi_{\nu}^{\boldsymbol{k}_\nu}(\boldsymbol{r})^* = \phi_{\mu}^{\boldsymbol{k}_\mu}(\boldsymbol{r}) \; \phi_{\nu}^{-\boldsymbol{k}_\nu}(\boldsymbol{r}) \\
    & \approx
    \sum_I \xi_I^{\boldsymbol{q}}(\boldsymbol{r}) \;
    X_{I\mu}^{\boldsymbol{k}_\mu} \; X_{I\nu}^{-\boldsymbol{k}_\nu}
    \label{eq:periodic-k-point-isdf}
    \end{aligned}
\end{equation}
where in the above equation, $X_{I\mu}^{\boldsymbol{k}_\mu} = \phi_{\mu}^{\boldsymbol{k}_\mu}(\boldsymbol{r}_I)$ is the interpolating vector (IV); 
$\boldsymbol{q} = \boldsymbol{k}_\mu - \boldsymbol{k}_\nu + \boldsymbol{B}_n$ due to momentum
conservation and is shifted to be within the first Brillouin zone ($\boldsymbol{B}_n$ is an integer multiple of the reciprocal lattice vectors); and $\xi_I^{\boldsymbol{q}}(\boldsymbol{r})$ is the interpolation function. 
The interpolation function is obtained from solving the least-squares problem corresponding to Eq.~\ref{eq:periodic-k-point-isdf}, which leads to the following linear equation,
\begin{equation}
\label{eq:periodic-k-point-isdf-linear-equation}
    \boldsymbol{\Pi}^{\boldsymbol{q}} \; \boldsymbol{\xi}^{\boldsymbol{q}} 
    = \boldsymbol{\eta}^{\boldsymbol{q}}
\end{equation}
where $\boldsymbol{\Pi}^{\boldsymbol{q}}$ is the metric tensor, defined from the product of IVs,
\begin{equation}
    {\Pi}^{\boldsymbol{q}}_{IJ} = \sum_{\boldsymbol{k}} \sum_{\mu\nu}
    X_{I\mu}^{\boldsymbol{k} - \boldsymbol{q}} \;
    X_{I\nu}^{-\boldsymbol{k}} \;
    X_{J\mu}^{\boldsymbol{q} - \boldsymbol{k}} \;
    X_{J\nu}^{\boldsymbol{k}}
\label{eq:periodic-k-point-isdf-metric-tensor}
\end{equation}
and which can be efficiently evaluated using the convolution theorem. 
Similarly, the right-hand side vector is evaluated as
\begin{equation}
    {\eta}^{\boldsymbol{q}}_I(\boldsymbol{r}) = \sum_{\boldsymbol{k}} \sum_{\mu\nu}
    X_{I\mu}^{\boldsymbol{k} - \boldsymbol{q}} \;
    X_{I\nu}^{-\boldsymbol{k}} \;
    \phi_{\mu}^{\boldsymbol{q} - \boldsymbol{k}}(\boldsymbol{r}) \;
    \phi_{\nu}^{\boldsymbol{k}}(\boldsymbol{r})
    \label{eq:periodic-k-point-isdf-right-hand-side-vector}
\end{equation}
The complete algorithm for both the metric tensor and right-hand side vector is outlined in Algorithm~\ref{alg:metric-tensor-and-right-hand-side-vector}: 
when evaluating the metric tensor, $\mathbf{Y}^{\boldsymbol{k}}$ are IVs, where index $f$ represents IPs and index $\mu$ corresponds to the AOs, while 
for the right-hand side vector, $\mathbf{Y}^{\boldsymbol{k}}$ represents AO values evaluated on the uniform grid, with index $f$ now labelling the grid points $\mathbf{r}$.

\begin{algorithm}[H]
    \caption{Efficient evaluation of metric tensor and right-hand side vector using convolution theorem.
    }
    \label{alg:metric-tensor-and-right-hand-side-vector}
    \setstretch{1.3}
    \begin{algorithmic}[1]
    \State \textbf{Input:}$X_{I\mu}^{\boldsymbol{k}}$, $Y_{f\mu}^{\boldsymbol{k}}$
    \State \textbf{Output:} $Z_{If}^{\boldsymbol{q}}$
    \For{each $\boldsymbol{k}$ in the Brillouin zone}
        \State $T_{If}^{\boldsymbol{k}} = \sum_\mu X_{I\mu}^{\boldsymbol{k}} Y_{f\mu}^{-\boldsymbol{k}}$
    \EndFor
    \State $T_{If}^{\boldsymbol{R}} \gets \text{FFT}^{-1}[T_{If}^{\boldsymbol{k}}]$ \Comment{transform to the supercell space}
    \State $Z_{If}^{\boldsymbol{R}} = T_{If}^{\boldsymbol{R}} \times T_{If}^{\boldsymbol{R}}$ \Comment{element-wise multiplication}
    \State $Z_{If}^{\boldsymbol{q}} \gets \text{FFT}[Z_{If}^{\boldsymbol{R}}]$ \Comment{transform back to the k-space}
    \end{algorithmic}
\end{algorithm}

After solving Eq.~\ref{eq:periodic-k-point-isdf-linear-equation} for the interpolation functions, their Coulomb kernel is defined as 
\begin{equation}
    W_{IJ}^{\boldsymbol{q}} = \int \text{d}\boldsymbol{r}_1 \text{d}\boldsymbol{r}_2 \;
    \frac{\xi_I^{-\boldsymbol{q}}(\boldsymbol{r}_1) \; \xi_J^{\boldsymbol{q}}(\boldsymbol{r}_2)}
    {|\boldsymbol{r}_1 - \boldsymbol{r}_2|}
    \label{eq:periodic-k-point-isdf-coulomb-matri-xi}
\end{equation}
which can be efficiently computed by FFT. 
Since the interpolation functions are themselves not required and only the Coulomb kernel enters into subsequent calculations, we can in fact determine the Coulomb kernel directly. This can be done by first computing $\mathbf{V}^{\boldsymbol{q}}$
\begin{equation}
    V_{IJ}^{\boldsymbol{q}} = \int \text{d}\boldsymbol{r}_1 \text{d}\boldsymbol{r}_2 \;
    \frac{\eta^{-\boldsymbol{q}}_I(\boldsymbol{r}_1) \; \eta^{\boldsymbol{q}}_J(\boldsymbol{r}_2)}
    {|\boldsymbol{r}_1 - \boldsymbol{r}_2|}
    \label{eq:periodic-k-point-isdf-coulomb-matri-eta}
\end{equation}
and then obtaining the Coulomb kernel from solving the least-squares linear equation,
\begin{equation}
    \boldsymbol{\Pi}^{\boldsymbol{q}} \; \mathbf{W}^{\boldsymbol{q}} \; \boldsymbol{\Pi}^{\boldsymbol{q}} 
    = \mathbf{V}^{\boldsymbol{q}}
    \label{eq:periodic-k-point-isdf-coulomb-matri-eta-trick}
\end{equation}

Once the ISDF Coulomb kernel is available, the two-electron integrals in the crystalline basis can be expressed in the THC factorized form,
\begin{equation}
    (\mu\boldsymbol{k}_\mu \nu\boldsymbol{k}_\nu 
    | \lambda\boldsymbol{k}_\lambda \sigma\boldsymbol{k}_\sigma) 
    = \sum_{IJ} 
    X_{I\mu}^{\boldsymbol{k}_\mu}
    X_{I\nu}^{-\boldsymbol{k}_\nu}
    X_{J\lambda}^{\boldsymbol{k}_\lambda}
    X_{J\sigma}^{-\boldsymbol{k}_\sigma}
    W_{IJ}^{\boldsymbol{q}}
    \label{eq:periodic-k-point-isdf-two-electron-integral}
\end{equation}

\subsection{Implementation of quantum embedding and local correlation with FFTISDF}

Starting from the two-electron integrals in the form of Eq.~\ref{eq:periodic-k-point-isdf-two-electron-integral}, the various components of quantum embedding and local correlation calculations can be accelerated. 

For example, the starting point is typically a periodic Hartree-Fock calculation.
The factorized Hamiltonian leads to an efficient evaluation of the exact exchange matrix~\cite{rettig-even-2023,yeh-lowscaling-2023,yeh-lowscaling-2024}
for the k-point Hartree-Fock (k-HF) theory,
\begin{equation}
\begin{aligned}
    K_{\mu \nu}^{\boldsymbol{k}} & = \frac{-1}{N_{\text{k}}}
    \sum_{IJ} \sum_{\boldsymbol{q}} X_{I\mu}^{\boldsymbol{k}} \, 
    X_{J\nu}^{-\boldsymbol{k}} \, W_{IJ}^{\boldsymbol{q}} 
    \sum_{\lambda \sigma}  X_{I\lambda}^{\boldsymbol{q} - \boldsymbol{k}}\, 
    P_{\lambda \sigma}^{\boldsymbol{k} - \boldsymbol{q}} \, 
    X_{J\sigma}^{\boldsymbol{k} - \boldsymbol{q}} \\
    & = \frac{-1}{N_{\text{k}}} \sum_{IJ} X_{I\mu}^{\boldsymbol{k}} \, 
    X_{J\nu}^{-\boldsymbol{k}} \, \sum_{\boldsymbol{q}} W_{IJ}^{\boldsymbol{q}} 
    \rho_{IJ}^{\boldsymbol{k} - \boldsymbol{q}}
\end{aligned}
\label{eq:periodic-k-point-isdf-exact-exchange-matrix}
\end{equation}
which can also be computed as a convolution.

Similarly, in local correlation frameworks it is generally necessary to do a global (i.e. on the whole system) correlation calculation at a low level of theory (e.g. second order perturbation theory or direct random phase approximation), either to determine the virtual correlation space, or to compute a global energy correction~\cite{aryasetiawan-2004,chang-2024,scott-2024,yeh-lowscaling-2024,song-2025}. 
The k-point FFTISDF formulation reduces the k-point scaling of such calculations also. In the case of k-point scaled-opposite-spin MP2 (k-SOS-MP2) and the 
k-point direct random phase approximation (k-dRPA) one can obtain overall linear scaling with the number of k-points~\cite{yeh-lowscaling-2023,yeh-lowscaling-2024}.

The k-SOS-MP2 correlation energy is defined by multiplying an empirical scaling factor $c_{\text{SOS}} = 1.3$ to the 
opposite-spin MP2 correlation energy,
\begin{equation}
\label{eq:periodic-k-point-isdf-sosp-mp2-energy}
    E^{\text{SOS\text{-}MP2}}_{\text{corr}} = c_{\text{SOS}} \, E_{\text{OS}},
\end{equation}
For an efficient formulation, the opposite-spin contribution can be computed via 
the imaginary-frequency non-interacting density response function~\cite{dixit-improving-2016} (the same expression can be derived from the Laplace-transformed energy 
denominator formula~\cite{song-atomic-2016,haritan-efficient-2025}),
\begin{equation}
\begin{aligned}
\label{eq:periodic-k-point-isdf-sos-mp2-os}
    E_{\text{OS}} &= \sum_{\boldsymbol{q}} 
    \int_{-\infty}^{\infty} \frac{\text{d} \omega}{2 \pi N_{\text{k}}} \; 
    \tr \Big[ \mathbf{P}^{\boldsymbol{q}}(i\omega) \, 
    \mathbf{P}^{\boldsymbol{q}}(i\omega)
    \Big]\\
    &= \sum_{\boldsymbol{q}} 
    \int_{0}^{\infty} \frac{\text{d} \tau}{N_{\text{k}}} \; 
    \tr \Big[ \mathbf{P}^{\boldsymbol{q}}(\tau) \, 
    \mathbf{P}^{\boldsymbol{q}}( - \tau)
    \Big]
\end{aligned}
\end{equation}
$\mathbf{P}^{\boldsymbol{q}}(\tau)$ 
can be efficiently evaluated with the convolution theorem,
using a similar algorithm as in Algorithm~\ref{alg:metric-tensor-and-right-hand-side-vector},
\begin{equation}
    P_{IJ}^{\boldsymbol{q}}(\tau) = \sum_{\boldsymbol{k}} \sum_{K}
    G_{IK}^{\boldsymbol{k}}(\tau) \;
    G_{IK}^{\boldsymbol{q} - \boldsymbol{k}}( - \tau) \; W_{KJ}^{\boldsymbol{q}}
\end{equation}
where $G_{IJ}^{\boldsymbol{k}}(\tau)$ is the imaginary-time non-interacting 
Green's function in the interpolating point representation,
\begin{equation}
    G_{IJ}^{\boldsymbol{k}}(\tau) = \sum_{\mu\nu} X_{I\mu}^{\boldsymbol{k}} \; X_{J\nu}^{-\boldsymbol{k}} \; 
    G_{\mu \nu}^{\boldsymbol{k}}(\tau)
\end{equation}
and in the atomic orbital basis, the Green's function is given by ($\tau > 0$),
\begin{equation}
\begin{aligned}
    G_{\mu \nu}^{\boldsymbol{k}}(+\tau) & = + &\sum_{i\in\text{occ}}
    C_{\mu i}^{\boldsymbol{k}} \; C_{i\nu}^{\boldsymbol{k}} 
    \exp(- \tau \epsilon^{\boldsymbol{k}}_i)
    \\
    G_{\mu \nu}^{\boldsymbol{k}}(-\tau) & = - &\sum_{a\in\text{vir}}
    C_{\mu a}^{\boldsymbol{k}} \; C_{a\nu}^{\boldsymbol{k}} \exp(+ \tau \epsilon^{\boldsymbol{k}}_a)
\end{aligned}
\end{equation}
$C_{\mu i}^{\boldsymbol{k}}$ and $\epsilon^{\boldsymbol{k}}_i$ 
($C_{\mu a}^{\boldsymbol{k}}$ and $\epsilon^{\boldsymbol{k}}_a$) are the 
k-point occupied (virtual) coefficients and energies obtained from a k-HF 
calculation, respectively.

For the k-dRPA, the correlation energy is
\begin{equation}
\label{eq:periodic-k-point-isdf-k-rpa-energy}
E^{\text{dRPA}}_{\text{corr}}
= \sum_{\boldsymbol{q}} 
\int_{-\infty}^{\infty} 
\frac{\text{d}\omega}{2\pi N_{\text{k}}}\,
\Big[\, \ln \det \mathbf{Q}^{\boldsymbol{q}}(\omega) 
+ \tr\, \mathbf{P}^{\boldsymbol{q}}(\omega) \Big]
\end{equation}
where $\mathbf{Q}^{\boldsymbol{q}}(\omega)=\mathbf{1}-\mathbf{P}^{\boldsymbol{q}}(\omega)$,
$\mathbf{1}$ is the identity matrix of dimension $N_{\text{IP}} \times N_{\text{IP}}$.
Eq.~\ref{eq:periodic-k-point-isdf-sos-mp2-os} is obtained as the second-order term in the 
Taylor expansion of the logarithm in Eq.~\ref{eq:periodic-k-point-isdf-k-rpa-energy}.
The integrals in Eqs.~\ref{eq:periodic-k-point-isdf-sos-mp2-os} and \ref{eq:periodic-k-point-isdf-k-rpa-energy} can be computed by numerical quadrature. 
Specifically, in this work we use modified Gauss-Legendre grids on the imaginary axis and 
Clenshaw-Curtis grids on the real axis.


A common component in both quantum embedding (such as density matrix embedding) and local natural orbital correlation treatments is the need to construct a fragment Hamiltonian in a subset of orbitals of the full problem. 
We can use the FFTISDF form of the Hamiltonian to accelerate  this construction. 
A similar formulation was previously proposed in Ref.~\cite{yeh-lowscaling-2024}.  
For a general description of the fragment Hamiltonians in density matrix embedding and local natural orbital correlation frameworks, we refer the reader to
Ref.~\cite{cui-efficient-2020} and Ref.~\cite{ye-periodic-2024}. 


The fragments are defined in terms of a set of fragment orbitals
\begin{equation}
\begin{aligned}
    \phi_{p}(\boldsymbol{r}) &= \frac{1}{\sqrt{N_{\text{k}}}} \sum_{\boldsymbol{k}} 
    \sum_{\mu} C_{\mu p}^{\boldsymbol{k}} \phi_{\mu}^{\boldsymbol{k}}(\boldsymbol{r}) 
\end{aligned}
\end{equation}
where the coefficients $C_{\mu p}^{\boldsymbol{k}}$ depend on the particulars of the embedding theory or correlation framework. {For example, in density matrix embedding, $\phi_p$ refers to the impurity and bath orbitals, while in local natural orbital correlation, they constitute the occupied and virtual orbitals of the fragment.} What is needed is the Hamiltonian expressed in the space spanned by the fragment orbitals $\{ \phi_p(\boldsymbol{r})\}$. 
The computational bottleneck lies in evaluating the two-body part of this Hamiltonian.
One way to obtain the two-body part is to transform from the k-space crystal orbital Hamiltonian to the fragment space,
\begin{equation}
    \begin{aligned}
        V_{pqrs} &= \frac{1}{N_{\text{k}}^2} \sum_{\boldsymbol{k}_\mu \boldsymbol{k}_\nu}
        \sum_{\mu \nu} 
        \sum_{\boldsymbol{k}_\lambda \boldsymbol{k}_\sigma} \sum_{\lambda \sigma}
        C_{\mu p}^{\boldsymbol{k}_\mu} C_{\nu q}^{-\boldsymbol{k}_\nu} 
        C_{\lambda r}^{\boldsymbol{k}_\lambda} C_{\sigma s}^{-\boldsymbol{k}_\sigma} \\
        & \quad \times (\mu\boldsymbol{k}_\mu \nu\boldsymbol{k}_\nu 
        | \lambda\boldsymbol{k}_\lambda \sigma\boldsymbol{k}_\sigma)
    \end{aligned}
\end{equation}
By employing the FFTISDF formula from Eq.~\ref{eq:periodic-k-point-isdf-two-electron-integral}, 
this transformation can be expressed in the more computationally efficient form:
\begin{equation}
\label{eq::vpqrs}
    V_{pqrs} = \sum_{\boldsymbol{q}} \sum_{IJ} 
    R_{I, pq}^{\boldsymbol{q}} \;
    R_{J, rs}^{-\boldsymbol{q}} \;
    W_{IJ}^{\boldsymbol{q}}
\end{equation}
where the intermediate tensor is,
\begin{equation}
    R_{I, pq}^{\boldsymbol{q}} = \frac{1}{\sqrt{N_{\text{k}}}} \sum_{\boldsymbol{k}}
    \sum_{\mu \nu} 
    X_{I\mu}^{\boldsymbol{k}} \; X_{I\nu}^{\boldsymbol{q}-\boldsymbol{k}}
    C_{\mu p}^{\boldsymbol{k}} \; C_{\nu q}^{\boldsymbol{q}-\boldsymbol{k}}
\end{equation}
The convolution theorem can be used to evaluate the intermediate tensor.
The procedure is summarized in Algorithm~\ref{alg:periodic-k-point-fftisdf}. 
Generating the Hamiltonian with FFTISDF has a cost scaling of 
$\mathcal{O}(N_{\mathrm{k}} N_{\mathrm{EO}}^2 N_{\mathrm{IP}}^2 + N_{\mathrm{k}} N_{\mathrm{EO}}^4 N_{\mathrm{IP}})$, which is more efficient than
the direct evaluation of $V_{pqrs}$ by FFTDF when
$N_{\mathrm{IP}}$ is much less than the total number of grid points.


\begin{algorithm}[H]
    \caption{Pseudocode for computing the two-body part of the embedding Hamiltonian.}
    \label{alg:periodic-k-point-fftisdf}
    \setstretch{1.3}
    \begin{algorithmic}[1]
    \State \textbf{Input:} $X_{I \mu}^{\boldsymbol{k}}$, $C_{\mu q}^{\boldsymbol{k}}$
    \State \textbf{Output:} $R_{I, pq}^{\boldsymbol{q}}$
    \For{all $\boldsymbol{k}$}
        \State $T_{I p}^{\boldsymbol{k}} = \sum_{\mu} X_{I \mu}^{\boldsymbol{k}} C_{\mu p}^{\boldsymbol{k}}$
    \EndFor
    \State $T_{I p}^{\boldsymbol{R}} \gets T_{I p}^{\boldsymbol{k}}$ \Comment{transform to supercell space}
    \State $R_{I, pq}^{\boldsymbol{R}} = T_{I p}^{\boldsymbol{R}} T_{I q}^{\boldsymbol{R}}$ \Comment{outer product}
    \State $R_{I, pq}^{\boldsymbol{k}} \gets R_{I, pq}^{\boldsymbol{R}}$ \Comment{transform to k-space}
    \end{algorithmic}
\end{algorithm}

\subsection{Parallelization}

The k-point FFTISDF formulation lends itself to parallelization, which we have implemented using MPI. Our approach begins by dividing the real-space grid 
into optimally sized chunks across available processors, which enables highly efficient parallel 
evaluation of the right-hand side vector following Eq.~\ref{eq:periodic-k-point-isdf-right-hand-side-vector}. 
We note that this domain decomposition significantly reduces the memory requirements per node while 
maintaining excellent load balancing across computational resources.
The Coulomb kernel $\mathbf{W}^{\boldsymbol{q}}$ is then computed in parallel for each k-point 
using Eqs.~\ref{eq:periodic-k-point-isdf-coulomb-matri-eta} and 
\ref{eq:periodic-k-point-isdf-coulomb-matri-eta-trick}.
For the numerical integrations 
required in k-SOS-MP2 and k-RPA calculations, we also distribute the integration points on 
the real or imaginary axes.

\subsection{Computational Details}
\label{sec:computational-details}

\begin{figure*}[t]
    \centering
    \includegraphics[width=1.0\textwidth]{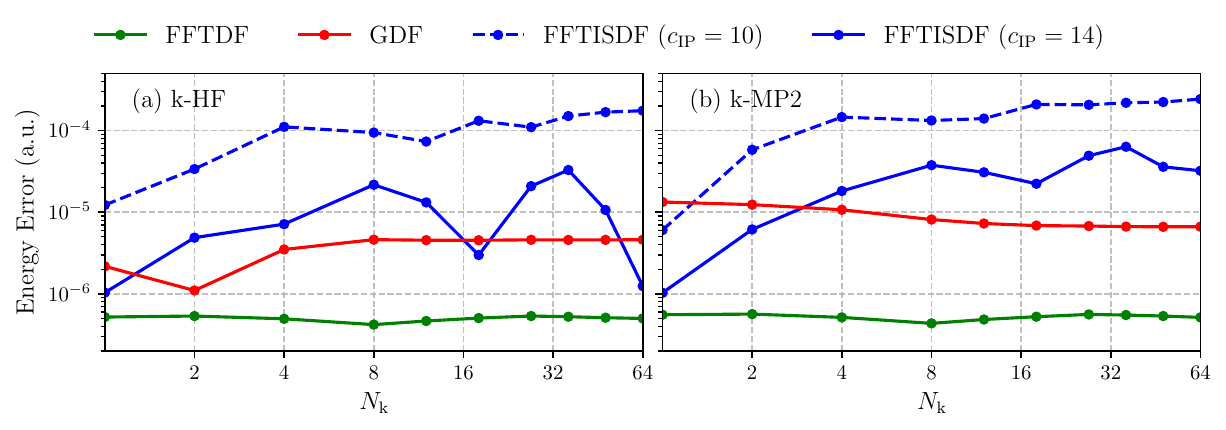}
    \caption{
        Per-atom energy errors for diamond obtained with: FFTDF ($E_{\text{cut}} = 60~\text{a.u.}$, green), GDF ($\beta = 2.0$, red), and FFTISDF ($E_{\text{cut}} = 60~\text{a.u.}$, blue). For FFTISDF, results are shown for $c_{\text{IP}} = 10$ (dashed) and $c_{\text{IP}} = 14$ (solid). Errors are computed with respect to a reference FFTDF calculation ($E_{\text{cut}} = 160~\text{a.u.}$). Subfigures correspond to (a) k-HF and (b) k-MP2.
    }
    \label{fig:diamond-krhf-and-krmp2}
\end{figure*}

We consider four representative solids with different electronic properties and structural characteristics: diamond as an example of a wide-bandgap semiconductor, carbon dioxide (CO$_2$) as a representative molecular crystal with weak intermolecular interactions, nickel monoxide (NiO) as a correlated transition metal oxide exhibiting Mott insulating behavior, and the infinite-layer CaCuO$_2$ (CCO) as a prototypical cuprate superconductor material with a layered perovskite structure.

All calculations were performed on Intel Icelake 8352Y processors (32-core, 2.2 GHz) with parallelization. 
For all systems, we employed correlation-consistent GTH-cc-pVDZ Gaussian basis sets~\cite{ye-correlationconsistent-2022} and the Goedecker-Teter-Hutter family of pseudopotentials specifically optimized for Hartree-Fock calculations~\cite{goedecker-separable-1996,hartwigsen-relativistic-1998,hutter-gth-2024}.
For Brillouin zone sampling, $\Gamma$-centered k-point meshes were utilized, and all mean-field calculations were converged to an accuracy of $10^{-6}$ a.u. per unit cell.
To address the integrable divergence of exact exchange, we applied the leading-order exchange 
finite-size correction (probe-charge Ewald, \texttt{exxdiv=ewald} in \textsc{PySCF})~\cite{sundararaman-regularization-2013,sun-gaussian-2017}.
For the correlated wavefunction solver, we used coupled cluster singles and doubles (CCSD) as well as with perturbative triples (CCSD(T)) as implemented in \textsc{PySCF}~\cite{sun-pyscf-2018,sun-recent-2020}. 
The DMET calculations were conducted using the interacting bath approach with correlation-potential fitting, but without charge self-consistency (IB w/o CSC in Ref.~\cite{cui-efficient-2020}), 
utilizing a modified version of the \textsc{LibDMET} package.
The LNO-based local correlation calculations were carried out through a module implemented in \textsc{PySCF-forge}~\cite{ye-periodic-2024}, utilizing the cluster-specific bath natural orbitals (CBNO) scheme~\cite{nusspickel-effective-2023}.

To establish optimal FFTISDF parameters, we implemented a systematic convergence procedure.
Initially, reference Hartree-Fock calculations in the Gaussian-Plane-Wave framework were performed for a series of k-point meshes, progressively increasing the plane-wave cutoff energy $E_{\text{cut}}$ until the energy per atom (for all meshes) achieved convergence within a threshold of $\epsilon_1$. Subsequently, this converged $E_{\text{cut}}$ value was used as the standard for both FFTDF and FFTISDF calculations.
Building upon this, the FFTISDF calculations employ the established $E_{\text{cut}}$, whereas the interpolation point scaling factor $c_{\text{IP}}$ was systematically adjusted to maintain accuracy relative to the FFTDF reference within a threshold of $\epsilon_2$ for both k-HF and k-MP2.
Additionally, Gaussian density fitting (GDF) was used as an alternative reference method, implemented with even-tempered auxiliary basis sets using an exponent parameter $\beta = 2.0$. 
Meanwhile, additional parameters specific to the DMET and LNO calculations were established following recommendations from Ref.~\cite{cui-efficient-2020} and Ref.~\cite{ye-periodic-2024}.

The optimized parameters for each system are  summarized in Table~\ref{tab:computational-parameters}, with convergence thresholds set to $\epsilon_1 = 10^{-6}$ a.u. per atom and $\epsilon_2 = 10^{-4}$ a.u. per atom.

\begin{table}[h]
    \caption{Number of atoms, number of atomic orbitals and 
    chosen computational parameters for the GTH-cc-pVDZ basis.}
    \label{tab:computational-parameters}
    \centering
    \setlength{\tabcolsep}{6pt}
    \renewcommand{\arraystretch}{1.4}
    \begin{tabular}{lcccc}
        \toprule
        System & $N_{\text{atom}}$ & $N_{\text{AO}}$ & 
        $E_{\text{cut}}$ (a.u.)& $c_{\text{IP}}$ \\
        \midrule
        Diamond & 2  & 26  & 60.0  & 14.0 \\
        CO$_2$  & 12 & 156 & 140.0 & 14.0 \\
        NiO (AFM cell) & 4  & 78  & 180.0 & 25.0 \\
        CCO ($2 \times 2$ cell) & 16 & 308 & 180.0 & 25.0 \\
        \bottomrule
    \end{tabular}
\end{table}
\section{Results and Discussion}
\label{sec:results}
\subsection{Diamond}

\begin{figure*}[bt]
    \centering
    \includegraphics[width=1.0\textwidth]{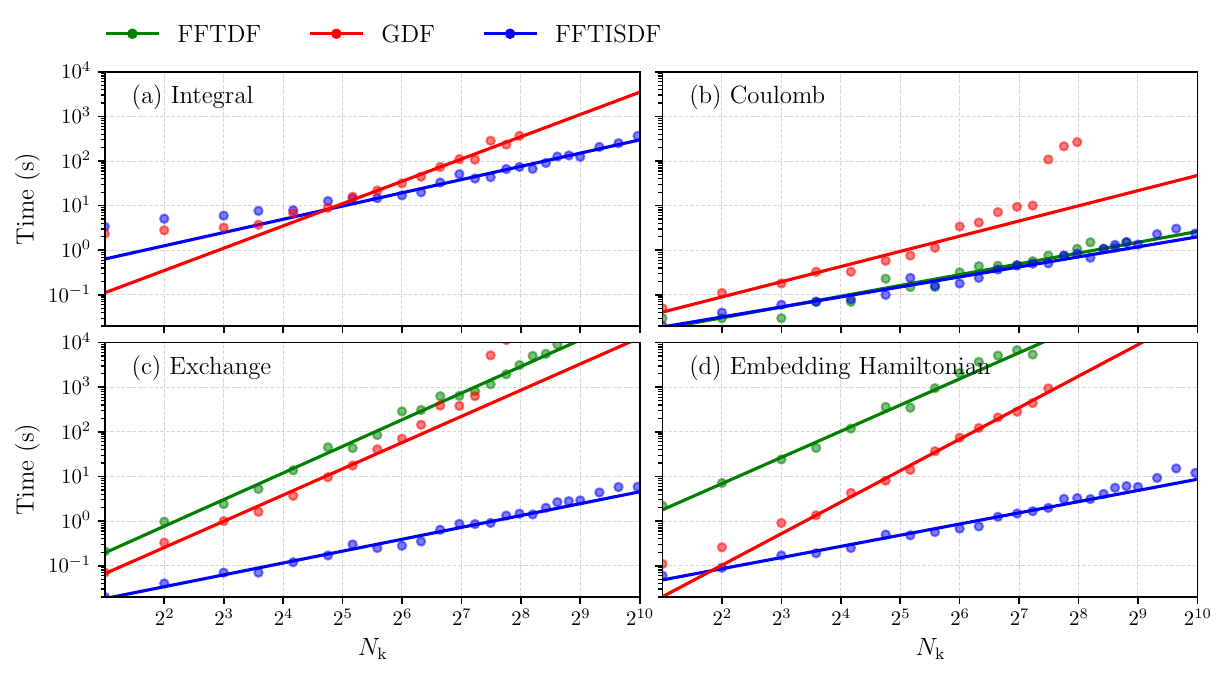}
    \caption{
    Comparison of wall-clock times for different density-fitting schemes applied to the diamond system, plotted against the number of k-points ($N_{\text{k}}$). Panels show the timing for: (a) integral evaluation, (b) Coulomb matrix construction, (c) exchange matrix construction, and (d) embedding Hamiltonian construction. Solid lines indicate fitted trends highlighting the computational scaling (linear or quadratic) of each method.
    }
    \label{fig:diamond-krhf-dmet-time}
\end{figure*}

\begin{table*}[t]
\caption{
Convergence of ground state and correlation energies (in a.u.\ per atom) for diamond with respect to k-point sampling.
The second column reports the k-HF energy, while columns 3--8 present correlation energies obtained from DMET (with CCSD solver), k-SOS-MP2, k-dRPA,  LNO-MP2, LNO-CCSD, and LNO-CCSD(T).
The LNO-based methods are extrapolated to the full correlation domain using k-SOS-MP2 as a baseline.
The bottom row shows values extrapolated to TDL.
}
    \centering
    \setlength{\tabcolsep}{1.8pt}
    \renewcommand{\arraystretch}{1.3}
    \begin{tabular*}{\textwidth}{@{\extracolsep{\fill}}cccccccc}
        \hline
         & $E$ (a.u.) & \multicolumn{6}{c}{$E_{\text{corr}}$ (a.u.)} \\
         \cline{3-8}
        $N_{\text{k}}$ & k-HF & DMET &  k-SOS-MP2 & k-dRPA  & LNO-MP2 & LNO-CCSD & LNO-CCSD(T) \\
\hline
$      2 \times 2 \times 2 $ &       -5.4786 &      -0.0814 &      -0.1112 &      -0.1269 &      -0.1174 &      -0.1237 &      -0.1292  \\
$      3 \times 3 \times 3 $ &       -5.5111 &      -0.0897 &      -0.1169 &      -0.1321 &      -0.1257 &      -0.1316 &      -0.1364  \\
$      4 \times 4 \times 4 $ &       -5.5144 &      -0.0938 &      -0.1196 &      -0.1342 &      -0.1299 &      -0.1343 &      -0.1395  \\
$      5 \times 5 \times 5 $ &       -5.5142 &      -0.0960 &      -0.1208 &      -0.1351 &      -0.1316 &      -0.1355 &      -0.1408  \\
$      6 \times 6 \times 6 $ &       -5.5137 &      -0.0975 &      -0.1213 &      -0.1355 &      -0.1323 &      -0.1361 &      -0.1414  \\
$      7 \times 7 \times 7 $ &       -5.5134 &      -0.0984 &      -0.1216 &      -0.1357 &      -0.1326 &      -0.1363 &      -0.1416  \\
$      8 \times 8 \times 8 $ &       -5.5131 &      -0.0992 &      -0.1217 &      -0.1358 &      -0.1328 &      -0.1364 &      -0.1418  \\
$   10 \times 10 \times 10 $ &       -5.5129 &      -0.1002 &      -0.1218 &      -0.1359 &          --- &          --- &          ---  \\
                         TDL &       -5.5127 &      -0.1010 &      -0.1219 &      -0.1361 &      -0.1331 &      -0.1367 &      -0.1421  \\
\hline
        \end{tabular*}
    \label{tab:diamond-tdl-extrapolation}
\end{table*}

We begin by assessing the accuracy and efficiency of FFTISDF using the diamond crystal as a benchmark. To ensure a rigorous comparison, we evaluated the ground-state energies (at both k-HF and k-MP2 levels) and the wall times for each computational step against established density fitting schemes (FFTDF and Gaussian Density Fitting (GDF)). This consistent setup allows for a direct evaluation of performance across all methods. To avoid Laplace-transform integration grid errors, the k-MP2 calculations with FFTISDF were obtained here from the exact MP2 expression using the FFTISDF two-electron integrals. The per-atom energy deviations are summarized in Fig.~\ref{fig:diamond-krhf-and-krmp2}, utilizing FFTDF with $E_{\text{cut}} = 160~\text{a.u.}$ as the reference. For FFTDF ($E_{\text{cut}} = 60~\text{a.u.}$), the error remains small ($< 10^{-6}~\text{a.u.}$ per atom), confirming the cutoff's sufficiency. Similarly, GDF yields consistent errors of approximately $10^{-5}~\text{a.u.}$ per atom, independent of the mesh size.

The FFTISDF {accuracy} clearly depends on both $c_\text{IP}$ and 
the number of k-points. With $c_{\text{IP}} = 10$, the error increases initially but approaches an asymptotic limit of approximately $10^{-3}$ a.u. for $N_{\text{k}} > 8$. 
The accuracy is systematically improved by increasing the interpolation point scaling factor. When $c_{\text{IP}} = 14$, the per-atom error is systematically reduced to below $10^{-4}~\text{a.u.}$. Consequently, we employ $E_{\text{cut}} = 60~\text{a.u.}$ and $c_{\text{IP}} = 14$ (as detailed in Table~\ref{tab:computational-parameters}) for all subsequent calculations.

Next, we evaluate the computational efficiency by analyzing the wall times required for computing integrals, Coulomb matrices, exchange matrices, and the fragment Hamiltonian. The results are plotted against the number of k-points in Fig.~\ref{fig:diamond-krhf-dmet-time}, with solid lines indicating the scaling behavior.
As shown in Fig.~\ref{fig:diamond-krhf-dmet-time}(a), the integral calculation time for GDF shifts from linear to quadratic scaling as $N_{\text{k}}$ increases. This change occurs as the computational bottleneck moves from the Cholesky decomposition of 2c2e integrals to the evaluation of 3c2e integrals. In contrast, FFTISDF maintains strict linear scaling across the entire range. While GDF is slightly faster at small $N_{\text{k}}$ (below $2^5$), FFTISDF outperforms it significantly as the system size grows. 
Note that FFTDF is excluded from this comparison as it avoids the computation of intermediate integrals.

For the Coulomb matrix evaluation [Fig.~\ref{fig:diamond-krhf-dmet-time}(b)], while all schemes scale linearly, GDF incurs a notably higher computational cost compared to FFTDF and FFTISDF. For exchange matrices [Fig.~\ref{fig:diamond-krhf-dmet-time}(c)], despite avoiding intermediate integral computations, FFTDF exhibits quadratic scaling and remains the most computationally expensive method. GDF also displays quadratic scaling, whereas FFTISDF is the only method that preserves linear scaling, demonstrating superior efficiency.
This advantage extends to the construction of the fragment Hamiltonian [Fig.~\ref{fig:diamond-krhf-dmet-time}(d)]. As FFTDF and GDF scale quadratically, they rapidly become computationally prohibitive. In contrast, FFTISDF preserves linear scaling, offering a speedup of several orders of magnitude at large $N_{\text{k}}$ and enabling efficient quantum embedding and local correlation calculations for up to 1000 k-points.

Beyond computational time, FFTISDF also offers superior storage efficiency to GDF, scaling as $N_k N_{\text{IP}}^2 + N_k N_{\text{IP}} N_{\text{AO}}$, in contrast to $N_k^2 N_{\text{AO}}^2 N_{\text{aux}}$ for GDF. This improvement results in a significant reduction in storage requirements, from 881.09 GB to 2.11 GB in the 1000 k-point diamond case. 

\begin{figure*}[t]
    \centering    
    \includegraphics[width=1.0\textwidth]{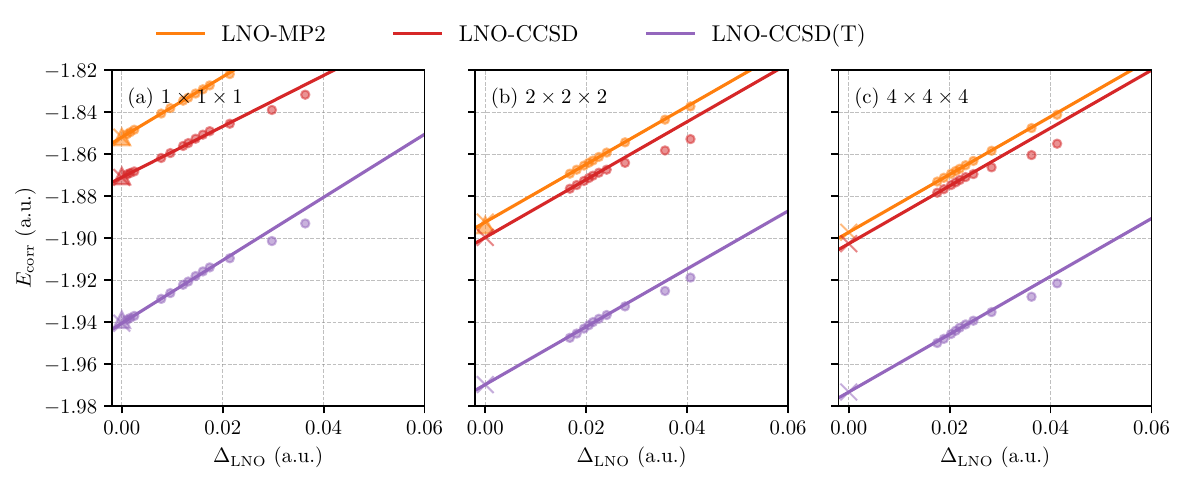}
    \caption{
        Extrapolation of CO$_2$ correlation energies. LNO-MP2, LNO-CCSD, and LNO-CCSD(T) correlation energies plotted against $\Delta_{\mathrm{LNO}}$ (the difference between k-SOS-MP2 and LNO-SOS-MP2 correlation energies) for k-point meshes: (a) $1 \times 1 \times 1$ ($\Gamma$-point only), (b) $2 \times 2 \times 2$, and (c) $4 \times 4 \times 4$. 
        Small circles represent LNO correlation energies, triangles show global k-point counterparts, 
        crosses mark extrapolated values, and solid lines are fitted lines.
    }
    \label{fig:co2-extrapolate}
\end{figure*}

\begin{figure}[t!]
    \includegraphics[width=0.48\textwidth]{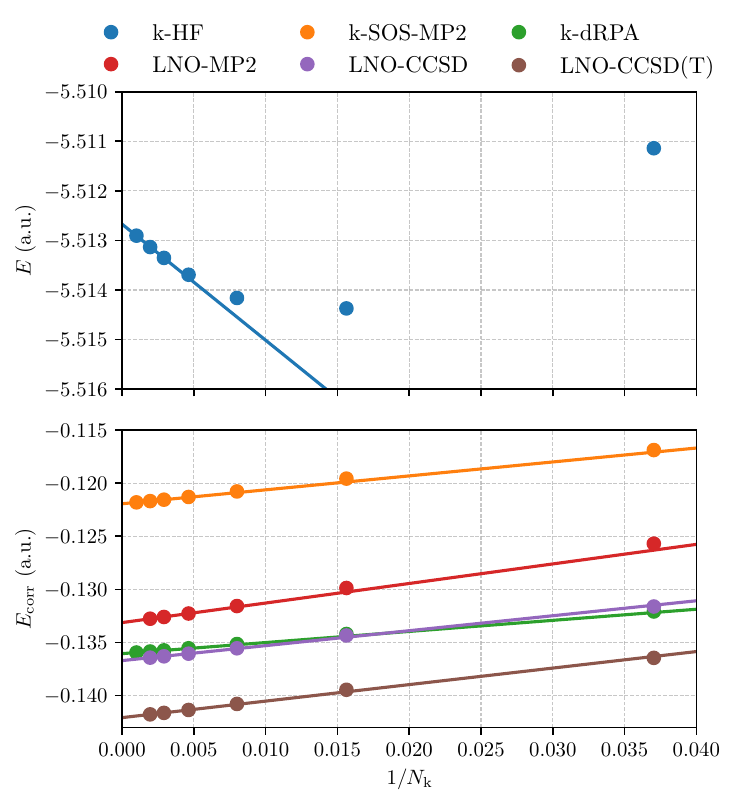}
    \caption{
        Thermodynamic limit extrapolation for diamond. Upper panel: k-HF total energy as a function of $1/N_{\text{k}}$. Lower panel: correlation energies from various methods including k-SOS-MP2, LNO-MP2, k-dRPA, LNO-CCSD, and LNO-CCSD(T). Solid lines represent linear fits used for extrapolation to the thermodynamic limit ($1/N_{\text{k}} \to 0$).
    }
    \label{fig:tdl}
\end{figure}

We next extrapolate the FFTISDF results to the thermodynamic limit. Table~\ref{tab:diamond-tdl-extrapolation} presents the per-atom k-HF energies and correlation energies from DMET, k-SOS-MP2, LNO-MP2, k-dRPA, LNO-CCSD, and LNO-CCSD(T) calculated for a series of cubic k-point meshes. To reduce the computational cost of LNO calculations for large correlation domains, we used k-SOS-MP2 energies to extrapolate the LNO-CCSD {and LNO-CCSD(T)} energies (details provided in the following subsection). The thermodynamic limit is then obtained by performing a linear fit against $N_{\text{k}}^{-1}$ using data from the four largest k-point meshes, as the Ewald exchange divergence correction ensures $\mathcal{O}(N_{\text{k}}^{-1})$ convergence behavior. The extrapolation is visualized in Fig.~\ref{fig:tdl}.

Notably, as more k-points are systematically incorporated, the k-HF energy initially decreases and then gradually increases before finally approaching the TDL value. While previously reported~\cite{goldzak-accurate-2022}, this energy-increasing regime remains prohibitive to sample adequately using existing methods (FFTDF and GDF) due to their steep scaling. Consequently, TDL extrapolations based on such truncated data are inherently less reliable. For correlated methods, DMET substantially underestimates the correlation energy as a result of its restricted  embedding space size, whereas the other correlated approaches yield more consistent values. Finally, the efficiency of FFTISDF makes it possible to extrapolate to reliable CCSD(T) energies at the thermodynamic limit for diamond.

\subsection{Carbon dioxide}

We select the carbon dioxide system as a representative molecular crystal, and
to demonstrate the extrapolation procedure to the complete correlation domain.
Initially, k-HF calculations were performed using the exchange matrix defined by Eq.~\ref{eq:periodic-k-point-isdf-exact-exchange-matrix}. Subsequently, the k-SOS-MP2 correlation energy was evaluated using Eq.~\ref{eq:periodic-k-point-isdf-sosp-mp2-energy}, establishing the reference point for the extrapolation.

Following this, LNO computations were performed across a range of truncation thresholds to obtain SOS-MP2, MP2, CCSD, and CCSD(T) correlation energies for each domain size. The difference between k-SOS-MP2 and LNO-SOS-MP2 correlation energies is denoted as $\Delta_{\mathrm{LNO}}$.
The MP2, CCSD, and CCSD(T) energies from the four largest correlation domains were then linearly fitted as functions of the corresponding $\Delta_{\mathrm{LNO}}$ values.

Finally, by extrapolating to $\Delta_{\mathrm{LNO}} = 0$ (corresponding to the complete k-SOS-MP2 limit), the global MP2, CCSD, and CCSD(T) correlation energies were obtained. This framework enables efficient evaluation of global correlation energies without incurring the prohibitive computational costs associated with direct large-scale calculations.

To validate the accuracy of our extrapolation procedure, we analyze the relationship between the LNO-SOS-MP2 correlation energies and their LNO-MP2, LNO-CCSD, and LNO-CCSD(T) counterparts across varying correlation domain sizes. Fig.~\ref{fig:co2-extrapolate}(a) illustrates the results for a $1 \times 1 \times 1$ k-point mesh ($\Gamma$-point), where the small circles denote the LNO energies and triangles represent the corresponding global k-point reference values. As expected, as the correlation domain expands, all LNO correlation energies (circles) systematically converge toward their global counterparts (triangles), confirming the correct asymptotic behavior. Based on this trend, we performed the extrapolation, with both the fitted lines and the resulting extrapolated values (crosses) displayed in the figure. The LNO-MP2 data points exhibit near-perfect linear alignment, resulting in exceptional agreement between the extrapolated value ($E^{\text{LNO-MP2}}_{\text{corr}} = -1.8519$ a.u.) and the global reference ($E^{\text{k-MP2}}_{\text{corr}} = -1.8517$ a.u.). The corresponding per-atom error is $\sim 1 \times 10^{-5}$ a.u.

The CCSD and CCSD(T) data show slight deviations from the fitted line for the small domains. 
This leads to minor differences between the 
extrapolated values ($E^{\text{LNO-CCSD}}_{\text{corr}} = -1.8710$ a.u.; $E^{\text{LNO-CCSD(T)}}_{\text{corr}} = -1.9405$ a.u.) and their global references ($E^{\text{k-CCSD}}_{\text{corr}} = -1.8703$ a.u.; $E^{\text{k-CCSD(T)}}_{\text{corr}} = -1.9398$ a.u.). Nevertheless, these errors remain remarkably small, with per-atom errors of $\sim 6 \times 10^{-5}$ a.u. in both cases.

We subsequently extended the extrapolation procedure to larger k-point meshes, specifically the $2 \times 2 \times 2$ and $4 \times 4 \times 4$ configurations, as illustrated in Fig.~\ref{fig:co2-extrapolate}(b) and (c). Although reference data is limited to k-MP2 for the $2 \times 2 \times 2$ mesh and is unavailable for other cases due to prohibitive computational costs, the overall trends remain consistent with those observed at the $\Gamma$-point.  Notably, we observe a substantial gap between the LNO-SOS-MP2 energy (using the largest affordable domain) and the exact global k-SOS-MP2 reference. At the $2 \times 2 \times 2$ mesh, this deviation amounts to $\sim 1 \times 10^{-3}$ a.u. per atom ($E^{\text{k-SOS-MP2}}_{\text{corr}} = -1.7750$ a.u. and $E^{\text{LNO-SOS-MP2}}_{\text{corr}} = -1.7582$ a.u.). 
A similar discrepancy is found in the MP2 correlation energy ($\sim 2 \times 10^{-3}$ a.u. per atom; $E^{\text{k-MP2}}_{\text{corr}} = -1.8956$ a.u. and $E^{\text{LNO-MP2}}_{\text{corr}} = -1.8707$ a.u.). However, the extrapolation  reduces the error to $\sim 1 \times 10^{-4}$ a.u. per atom. 
This result highlights the insufficiency of the computationally accessible truncated domains and establishes the effectiveness of our extrapolation strategy. Although global correlated benchmarks are unavailable for the larger k-point meshes, the LNO-based methods continue to exhibit excellent linearity. Consequently, we anticipate high reliability for the extrapolated values derived from these linear extrapolations. 

\begin{table}[h]
    \caption{
        Extrapolated TDL energies for HF, MP2, CCSD, and CCSD(T) for CO$_2$.
    }
    \setlength{\tabcolsep}{10.0pt}
    \renewcommand{\arraystretch}{1.3}
    \centering
    \begin{tabular}{ccc}
    \hline
    Method & $E_{\mathrm{corr}}$ (a.u.) & $E$ (a.u.) \\
    \hline
            HF &          --- &    -148.0545 \\
           MP2 &      -1.8979 &    -149.9524 \\
          CCSD &      -1.9030 &    -149.9574 \\
       CCSD(T) &      -1.9748 &    -150.0293 \\
    \hline
    \end{tabular}
    \label{tab:tdl-energies}
\end{table}

\begin{table*}[t!]
    \caption{
        Ground-state energies (in a.u.) and Heisenberg exchange constants $J$ (in meV) for NiO and CCO in the AFM and FM states, calculated using k-HF and DMET {(CCSD solver)} with various k-point meshes. The thermodynamic limit (TDL) is obtained by extrapolating the k-HF energies and DMET correlation energies separately using a linear fit against $N_{\text{k}}^{-1}$ with the three largest k-point grids. $J$ represents the next-nearest-neighbor exchange constant $J_2$ for NiO and the nearest-neighbor exchange constant $J_1$ for CCO.
        }
    \centering
    \renewcommand{\arraystretch}{1.3}
    \begin{tabular*}{0.9\textwidth}{@{\extracolsep{\fill}}ccccccc}
    \hline
     & \multicolumn{2}{c}{$E_{\text{AFM}}$ (a.u.)} & \multicolumn{2}{c}{$E_{\text{FM}}$ (a.u.)} & \multicolumn{2}{c}{$J$ (meV)} \\
     \cline{2-3}\cline{4-5}\cline{6-7}
    $N_{\text{k}}$ & k-HF & DMET & k-HF & DMET & k-HF & DMET \\
    \hline
    NiO & & & & & &  \\
$  2 \times 2 \times 2 $ &     -366.7825 &    -367.3968 &    -366.7794 &    -367.3918 &  -14.10 & -22.88 \\
$  3 \times 3 \times 3 $ &     -366.7687 &    -367.3778 &    -366.7669 &    -367.3744 &   -8.12 & -15.75 \\
$  4 \times 4 \times 4 $ &     -366.7622 &    -367.3766 &    -366.7605 &    -367.3734 &   -7.69 & -14.75 \\
$  5 \times 5 \times 5 $ &     -366.7593 &    -367.3770 &    -366.7576 &    -367.3738 &   -7.63 & -14.09 \\
$  6 \times 6 \times 6 $ &     -366.7580 &    -367.3785 &    -366.7563 &    -367.3753 &   -7.63 & -14.43 \\
                     TDL &     -366.7562 &    -367.3787 &    -366.7545 &    -367.3756 &   -7.59 & -14.05 \\

    & & & & & & \\
    CCO & & & & & &  \\
$  2 \times 2 \times 1 $ &    -1055.4085 &   -1056.5605 &   -1055.4012 &   -1056.5342 &  -50.02 & -178.69 \\
$  4 \times 4 \times 2 $ &    -1053.5604 &   -1054.5908 &   -1053.5545 &   -1054.5663 &  -40.35 & -167.10 \\
$  6 \times 6 \times 4 $ &    -1050.5774 &   -1051.6032 &   -1050.5715 &   -1051.5783 &  -40.66 & -169.14 \\
                     TDL &    -1053.2964 &   -1054.3095 &   -1053.2907 &   -1054.2851 &  -38.97 & -165.44 \\
    \hline
    \end{tabular*}
    \label{tab:nickel-oxide-afm-and-fm}
\end{table*}

Based on the validation of the LNO-extrapolation procedure, we further proceeded to obtain the total energies in the 
TDL by employing multiple k-point meshes. 
The final values are summarized in Table~\ref{tab:tdl-energies}.

\subsection{Nickel monoxide and cuprate}
We apply the FFTISDF-DMET framework {with a CCSD solver} to two prototypical strongly correlated transition metal oxides: NiO and CaCuO$_2$ (CCO). Our analysis focuses on the energy difference between antiferromagnetic (AFM) and ferromagnetic (FM) orderings. Notably, the implementation of such large k-point meshes is made feasible specifically by the FFTISDF algorithm. This allows us to systematically study long-range correlation effects that are critical for accurately determining the magnetic energy gaps in the strongly correlated regime.

When NiO is cooled below its N\'{e}el temperature, it adopts an AFM phase characterized by staggered magnetization along the [111] direction, commonly known as the AFM-II phase. By mapping the energy difference between the AFM and FM states to a spin Hamiltonian, one can derive the nearest-neighbor $J_1$ and next-nearest-neighbor $J_2$ exchange constants~\cite{macenulty-optimization-2023}. Since experimental and theoretical evidence indicates that $J_1$ is negligible in NiO~\cite{casselman-rightangled-1960}, we neglect its contribution and focus exclusively on computing $J_2$, the dominant exchange interaction, from the difference in the FM and AFM energies.

Table~\ref{tab:nickel-oxide-afm-and-fm} presents the ground-state energies obtained from k-HF and DMET calculations across various k-point meshes, using spin-polarized initial guesses. To approach the thermodynamic limit, we extrapolated the k-HF energies and DMET correlation energies separately by performing a linear fit against $N_{\text{k}}^{-1}$ using data from the three largest k-point meshes. The extrapolation reveals distinct behaviors between the two methods. While k-HF systematically underestimates $J_2$, DMET yields a value of $-14.05$ meV, slightly underestimating the magnitude compared to the experimental range of $-19.8$ to $-17.0$~meV~\cite{dep.r.moreira-effect-2002}. This improvement over k-HF demonstrates the importance of incorporating correlation effects in describing the magnetic interactions of NiO.

Turning to CCO, the defining structural feature of this cuprate compound is the two-dimensional CuO$_2$ square lattice plane (formally [CuO$_2$]$^{2-}$). While these planes are typically separated by buffer layers in various high-$T_c$ families, we focus here on the infinite-layer calcium cuprate CaCuO$_2$. Due to the layered structure, we employ anisotropic k-point meshes with fewer points along the $c$-axis. The correlation effects in CCO are even more pronounced than in NiO. At the thermodynamic limit, DMET yields $J = -165.44$ meV, more than four times larger in magnitude than the k-HF value of $-38.97$ meV. This DMET result is in good agreement with experimental values of $-142$ and $-158$ meV obtained by fitting the spin-wave dispersion near the $\Gamma$ point~\cite{kan-2004,peng-2017}.
\section{Conclusion}
\label{sec:conclusion}
We have described an implementation of interpolative separable density fitting in the Gaussian-Plane-Wave framework (FFTISDF in \textsc{PySCF}) and its integration with quantum embedding and local correlation frameworks. 
This implementation establishes a robust computational framework capable of obtaining converged ground-state energies from DMET and LNO-based MP2, CCSD, and CCSD(T) calculations for crystals with simple unit cells in the thermodynamic limit.
While this efficient methodology opens new avenues for high-level correlated calculations in extended systems, several aspects of the proposed approach still warrant further improvement.

First, the current implementation requires a relatively large number of interpolation points, typically ten times the number of atomic orbitals (see Table~\ref{tab:computational-parameters}), to reach our desired accuracy. As a result, the determination of these interpolation points becomes computationally demanding, particularly for systems with large unit cells. Developing efficient protocols for generating compact interpolation points would therefore be highly desirable.

Second, the size of the real-space grid employed in the fast Fourier transforms remains large. This is a challenge for systems containing transition metals or those with 
large unit cells. 

Finally, in this work, LNO-based approaches employed the CBNO scheme~\cite{nusspickel-effective-2023} for constructing correlation domains, although this method does not exhibit an optimal convergence with respect to the domain size~\cite{ye-periodic-2024}. Exploring alternative correlating spaces is of interest. Furthermore, the developed embedding framework can be readily extended to other quantum embedding approaches, such as dynamical mean field theory ~\cite{zgid-dynamical-2011,zhu-efficient-2020}.

\section{Acknowledgements}
This work was supported by the U.S. Department of Energy, Office of Science, through Award No. DE-SC0018140. 
G.K.C. is a Simons Investigator in Physics. 
We are grateful to Zhi-hao Cui and Shuoxue Li for their assistance with the DMET calculations. 
We thank Rui Li, Xing Zhang, Ruiheng Song, and Yao Luo for insightful discussions.
The code and data for this work are available at \url{https://github.com/yangjunjie0320/fftisdf}.

\bibliography{bibliography.bib}
\end{document}